\documentclass[conference]{IEEEtran}

\usepackage[dvipdfmx]{graphicx}

\usepackage{epsfig}
\ifCLASSINFOpdf
\usepackage[utf8]{inputenc}
\usepackage[fleqn]{amsmath}
\usepackage[dvipdfm]{graphicx}
\usepackage{latexsym}
\usepackage[varg]{txfonts}
\usepackage{bm}
\usepackage{color}
\usepackage{graphicx}
\usepackage{cases}
\usepackage{amsmath,amssymb}
\usepackage{subfig}
\usepackage{subfigure}

\else
\fi
\usepackage[cmex10]{amsmath}
\begin{document}
%
\title{Estimation of Relationship between \\Stimulation Current and Force Exerted \\ during Isometric Contraction}

\author{\IEEEauthorblockN{Tomoya Kitamura}
\IEEEauthorblockA{Graduate School of Science and Engineering,\\Saitama University,\\
255, Shimo-ohkubo, Sakura-ku, Saitama, 338--8570, Japan\\
Email: t.kitamura.360@ms.saitama-u.ac.jp\\}
\and
\IEEEauthorblockN{Yuu Hasegawa}
\IEEEauthorblockA{Graduate School of Science and Engineering,\\Saitama University,\\
255, Shimo-ohkubo, Sakura-ku, Saitama, 338--8570, Japan\\
Email: y.hasegawa.470@ms.saitama-u.ac.jp\\}
\and
\IEEEauthorblockN{Sho Sakaino}
\IEEEauthorblockA{Graduate School of Science and Engineering,\\Saitama University,\\
255, Shimo-ohkubo, Sakura-ku, Saitama, 338--8570, Japan\\ JST PRESTO\\
Email: sakaino@mail.saitama-u.ac.jp\\}
\and
\IEEEauthorblockN{Toshiaki Tsuji}
\IEEEauthorblockA{Graduate School of Science and Engineering,\\Saitama University,\\
255, Shimo-ohkubo, Sakura-ku, Saitama, 338--8570, Japan\\
Email: tsuji@ees.saitama-u.ac.jp\\}

}

\maketitle

\begin{abstract}
In this study, we developed a method to estimate the relationship between stimulation current and volatility during isometric contraction. In functional electrical stimulation (FES), joints are driven by applying voltage to muscles. This technology has been used for a long time in the field of rehabilitation, and recently application oriented research has been reported. However, estimation of the relationship between stimulus value and exercise capacity has not been discussed to a great extent. Therefore, in this study, a human muscle model was estimated using the transfer function estimation method with fast Fourier transform. It was found that the relationship between stimulation current and force exerted could be expressed by a first-order lag system. In verification of the force estimate, the ability of the proposed model to estimate the exerted force under steady state response was found to be good.\\
\end{abstract}



%
\IEEEpeerreviewmaketitle

\section{Introduction}
\begin{figure}[tbp]
\centering
\includegraphics[width=60mm]{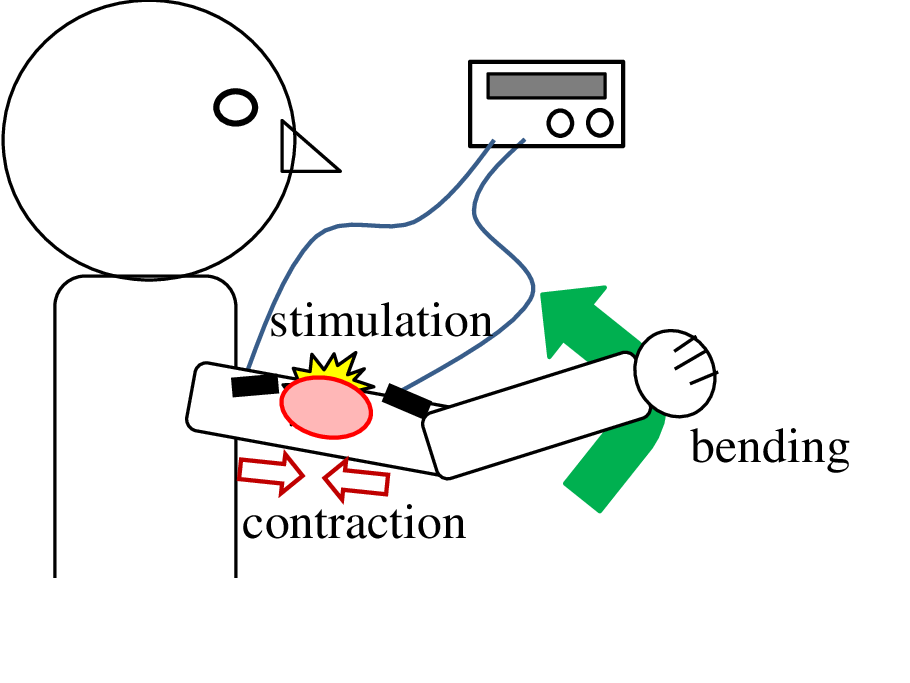}
\vspace{-5mm}
\caption{Conceptual diagram of FES}
\label{fig:fes2}
\end{figure}

In functional electrical stimulation (FES), a person's joints are driven by applying current to muscles \cite{c1} \cite{c2}. A schematic diagram of FES is shown in Fig.~\ref{fig:fes2}. In FES, muscle contraction is induced by applying a current to the muscle. The corresponding joint is driven accordingly. This technique has been used in the field of rehabilitation for a long time. Jaime {\it et al.} proposed a method of generating standing motion using FES by employing proportional-integral-derivative (PID) control \cite{c3}. Farhoud {\it et al.} proposed a pedaling operation based on an FES system composed of a sliding mode controller and a fuzzy mode controller \cite{c4}. Bouton {\it et al.} reported a method of generating desired actions read from electroencephalogram (EEG) by using FES \cite{c5}. 

In addition, research on application development for healthy subjects has been reported in recent years. Tamaki {\it et al.} reported hand control techniques based on FES \cite{c6}. Pedro et al. proposed a method for presenting the weight of virtual objects by using virtual reality (VR) goggles in conjunction with FES \cite{c7}. In addition, we proposed bilateral control based on FES \cite{c8} \cite{c9} \cite{c10} \cite{c22}. Bilateral control is a master slave system \cite{c19} \cite{c20} \cite{c21}. In bilateral control, the slave is controlled by FES to follow the movement of the master. In addition, the master can feel the reaction force at the slave. Therefore, the master can judge the state of the slave more clearly. 

In general, force control is required when controlling a robot or the like. This makes it possible to construct a good position control method. However, there are few reports on the relationship between stimulus value (i.e. applied voltage or inflow current) and force exerted. This is because, in recent years, methods of the relationship between stimulus value and angles value by neural network (NN) or the like is mainstream \cite{c11} \cite{c12} \cite{c13}. In body control for rehabilitation, the objective can be achieved if position control is possible. 

In extant research on force control, Ferrarin {\it et al.} proposed a method for approximating the relationship between the applied voltage and the force exerted by using a first-order lag system \cite{c14}. However, the basis for using the first-order lag system is not mentioned. Matsui {\it et al.} approximated the exiting torque from the inflow current by using a second-order lag system \cite{c15}. However, this method employs antagonistic muscle stimulation. Therefore, there is a possibility that characteristics different from those associated with simple muscle stimulation. In addition, the Hammerstein model \cite{c16}, and a few subsequent improvement have been reported \cite{c17} \cite{c18}. For control using FES, a method using voltage control \cite{c6} \cite{c8} \cite{c14} and a method using current control \cite{c4} \cite{c11} \cite{c12} have been mixed, but it was not reported as to which value determines the exerted force.

In this paper, we examine the relationship between stimulus value and exerted force. Firstly, the relationship between applied voltage and inflow current is investigated. Secondly, we confirm that the force is determined by the inflow current value. Thirdly, the inflow current is gradually increased, and the threshold current to drive the joint is measured. Next, the transfer function of the current and force are estimated using the transfer function estimation method with fast Fourier transform (FFT). As a result, we could confirm that the relationship between current and force can be represented by a first-order lag system with dead time. Finally, parameters are fitted by multiple regression analysis, and the proposed model is validated.

The remainder of this paper is organized as follows. First, Section II describes FES. Section III describes the transfer function estimation method using FFT. Section IV describes the experimental procedure and its results. Specifically, regarding estimation of the relationship between the voltage and the current, it is demonstrated that the force exerted is determined by inflow current, and the method for estimating  the relationship between the current and the force is described. Section V verifies the validity of the proposed method by using stimulus patterns that were not used for modeling. Section VI presents our concluding remarks. 

\section{Functional Electrical Stimulation}
\begin{figure}[tbh]
\centering
\includegraphics[width=40mm]{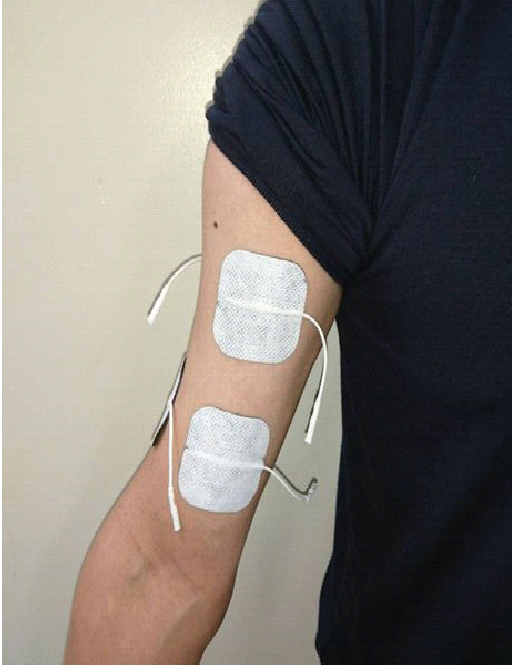}
    \caption{Stimulus location}
    \label{fig:kinnniku} 
\end{figure}

In this section, FES is described. In FES, the voltage is applied to an adhesive pad affixed to the skin surface to pass electric current into the muscle. Muscle contraction is caused by this current, and joints are driven. In this study, we stimulated the biceps brachii muscle and measured the force exerted by the elbow joint. Fig.~\ref{fig:kinnniku} shows the pad locations. In terms of the stimulation position, the motor point was specified using Compex Performance (``Compex, USA''), and an adhesive pad (dimensions: 50~$\times$~50~mm; AXELGAARD M MODEL 895220) was affixed at that position. 

Human body control by means of FES can be achieved using voltage control or current control. In general, voltage control is safer, because the resistance increases when the pad is about to come off, and the flow of current decreases. When using current control, the current is concentrated in a small area when the pad is about to come off, which is dangerous. In addition, with the flow of current, pH changes to acidic on the anode side and alkaline on the cathode side. With normal electrical stimulation, pH is maintained constant by blood flow, but if condition that pH is different between anode and cathode continues, injuries such as burns may occur. Therefore, we must be aware of the direction of current flowing.

The experimental machine that used in this paper is introduced. In this paper, control was performed by adjusting the voltage within the maximum value of 35 V. The circuit was configured to conform to the Japanese Industrial Standard (JIS). Therefore, the current flowing through the human body was not allowed to exceed 20 mA. 

\section{Estimation of Transfer Function using FFT}
In this section, a transfer function estimation method using FFT is described. The Fourier transform is expressed by the following equation:
\begin{equation}
U(f)=\int_{-\infty}^{\infty} u(t) e^{-jft} dt,
\label{eq:FT}  
\end{equation}
where $u(t)$ represents the time domain, and $U(f)$ represents the frequency domain. Discrete Fourier Transform (DFT) and FFT are generally used for Fourier transformation; in this paper, FFT, which can shorten calculation time, was adopted. The transfer function $ H (f) $ represents the relationship between the input $ X (f) $ and the output $ Y (f) $. In the frequency domain,
\begin{equation}
H(f)=\frac{Y(f)}{X(f)}.
\label{eq:TF}  
\end{equation}

However, noise is included in the actual measurement. Therefore, we rewrite (\ref{eq:TF}) as follows:
\begin{equation}
H(f)=\frac{Y(f) \cdot X(f)^{*}}{X(f) \cdot X(f)^{*}},
\label{eq:TF2}  
\end{equation}
where the superscript $*$ represents the complex conjugate of the complex spectrum obtained using the Fourier transform. The denominator in (\ref{eq:TF2}) is called the auto power spectrum, and the numerator is called the cross spectrum. By using (\ref{eq:TF2}), the influence of noise on the output side can be reduced. By confirming the gain margin and the phase margin of the obtained $ H (f) $, it is possible to estimate the transfer function.

\section{Experimental Methods and Results}
\begin{figure}[t]
\centering
\includegraphics[width=80mm]{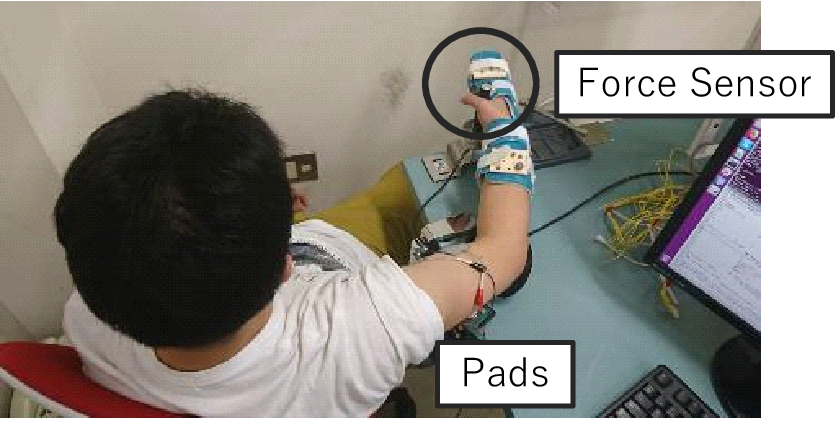}
    \caption{Scene of Experiment}
    \label{fig:View} 
\end{figure}

In this section, the procedures and the results of the experiment for modeling the current and the exerted force are described sequentially. Two healthy subjects (referred to as A and B) were used to test the proposed system. Informed consent was obtained from the participants, and the study was approved by Ethics Committee of  Saitama University. The experimental setup is indicated in Fig.~\ref{fig:View}. Subjects sat on the chair so that their arms were level with the floor. The hands of the subjects are fixed with Gibbs. Therefore, the wrist joint cannot exert a force. The subjects were made to allow the hands to hit the force sensor. The force sensor was used by PFS055YA251U6 (``Leptrino, Japan'').

\subsection{Relationship between Applied Voltage and Inflow Current}
\begin{figure}[t]
\centering
\includegraphics[width=80mm]{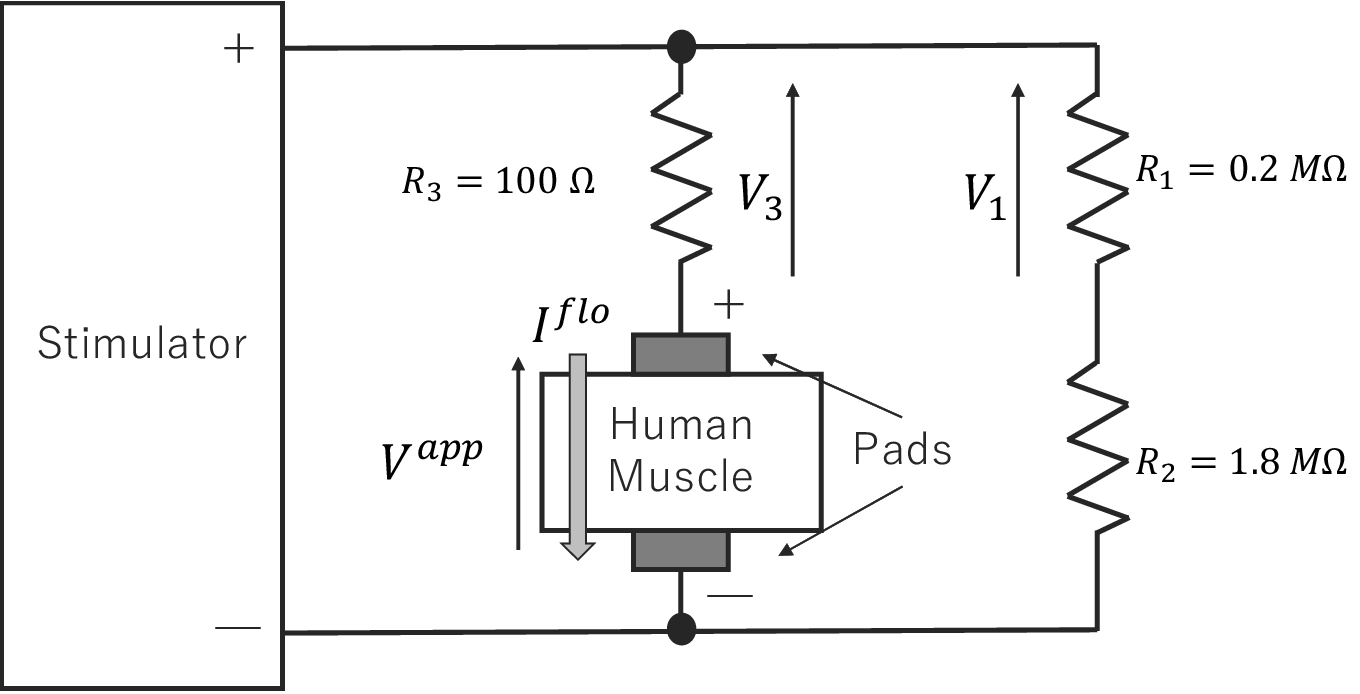}
    \caption{Placement of stimulator, pad, and resistor}
    \label{fig:Place} 
\end{figure}

In the experiment conducted herein, the stimulators, pads, and resistors were placed as shown in Fig.~\ref{fig:Place}. Assuming the voltage applied to a person is $V^{app}$ and the current flowing in the body is $I^{flo}$, the relationship with the voltage $V_{n}$ applied across the resistor $R_{n}$ can be written as follows:
\begin{eqnarray}
&V^{app}=V_{1}\cdot \frac{R_{1}+R_{2}}{R_{1}}-V_{3} \nonumber \\
&I^{flo}=\frac{V_{3}}{R_{3}}.
\label{eq:Circuit}  
\end{eqnarray}

In this paper, the resistances $R_{1} = 0.2 M~\Omega$, $R_{2} = 1.8 M~\Omega$, and $R_{3} = 100~\Omega$ were used. Therefore, (\ref{eq:Circuit}) can be rewritten as follows:
\begin{eqnarray}
V^{app}=10V_{1}-V_{3} \nonumber \\
I^{flo}=\frac{V_{3}}{100}.
\label{eq:Circuit2}  
\end{eqnarray}

\begin{figure}[t]
\centering
\includegraphics[width=80mm]{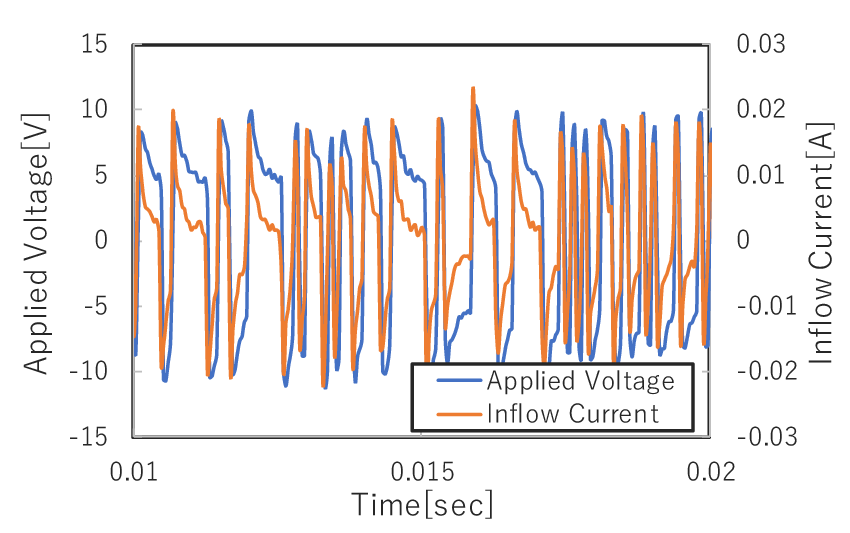}
    \caption{Applied voltage and inflow current of subject A}
    \label{fig:VI} 
\end{figure}

\begin{figure}[t]
\centering
\includegraphics[width=80mm]{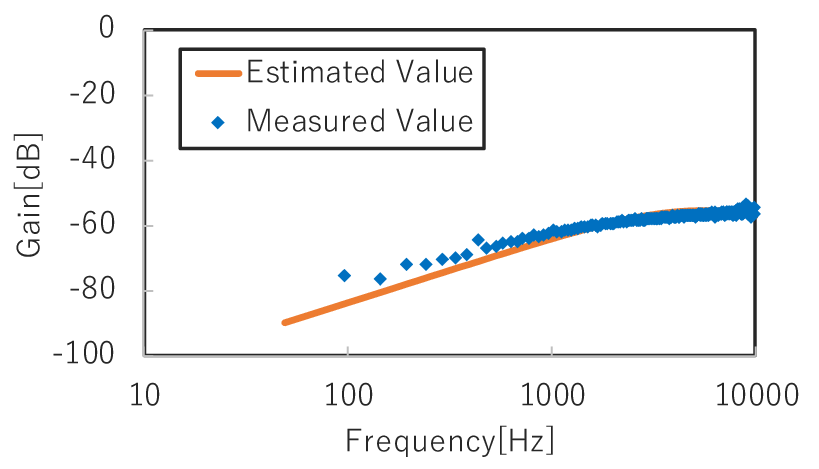}
\caption{Gain margin of current and voltage of subject A}
\label{fig:GVI1}
\end{figure}

We measured $V_ {1}$ and $V_ {3}$ by using an oscilloscope with a sampling period of $1.0$ $\mu$sec. Therefore, the sampling frequency was 1.0 MHz. The applied voltage $V ^ {app}$ and the actual inflow current $I ^ {flo}$ in this experiment are shown in Fig.~\ref{fig:VI}. As shown in Fig.~\ref{fig:VI}, the input voltage is an M-series signal that changes randomly at +10 or -10 V at a carrier frequency of 10 kHz. We calculated the relationship between the applied voltage and the inflow current using multiple regression analysis. As a result, the mathematical formula represented by the (\ref{eq:TF_VI}) has the highest correlation value.
\begin{equation}
 G_ {VI} (s)=\frac{I(s)}{V(s)}=\frac{b_{1}s}{a_{2}s^2+a_{1}s+1},
\label{eq:TF_VI}  
\end{equation}
where, $a_{2}, a_{1}$, and $b_{1}$ are constants. These three values were estimated using multiple regression analysis. The fixed values and resonance frequency of each subject are listed in Table~\ref{tb:Value1}. In addition, the gain margin in the subject A is shown in Fig.~\ref{fig:GVI1}. The results showed that the resonance frequency of $G_{VI} (s)$ is high frequency. Therefore, $G_{VI} (s)$ is represented by one differentiation in the low frequency range.

\begin{table}[tbh]
\centering
\caption{Values of parameter concerning $G_{VI} (s)$ and resonance frequency}
\label{tb:Value1}
  \begin{tabular}{|c||c|c|} \hline
     & Subject A & Subject B \\ \hline\hline
    $a_{2}$ & $8.0\times 10^{-10}$ & $1.2\times 10^{-9}$  \\ \hline
    $a_{1}$ & $2.2\times 10^{-5}$ & $4.6\times 10^{-5}$ \\ \hline
    $b_{1}$ & $1.9\times 10^{-7}$ & $1.4\times 10^{-7}$ \\ \hline\hline
Resonance frequency [kHz] & 5.6 & 4.6 \\ \hline
  \end{tabular}
\end{table}

\subsection{Stimulus with Long Pulse Width}
\begin{figure}[t]
\centering
\includegraphics[width=80mm]{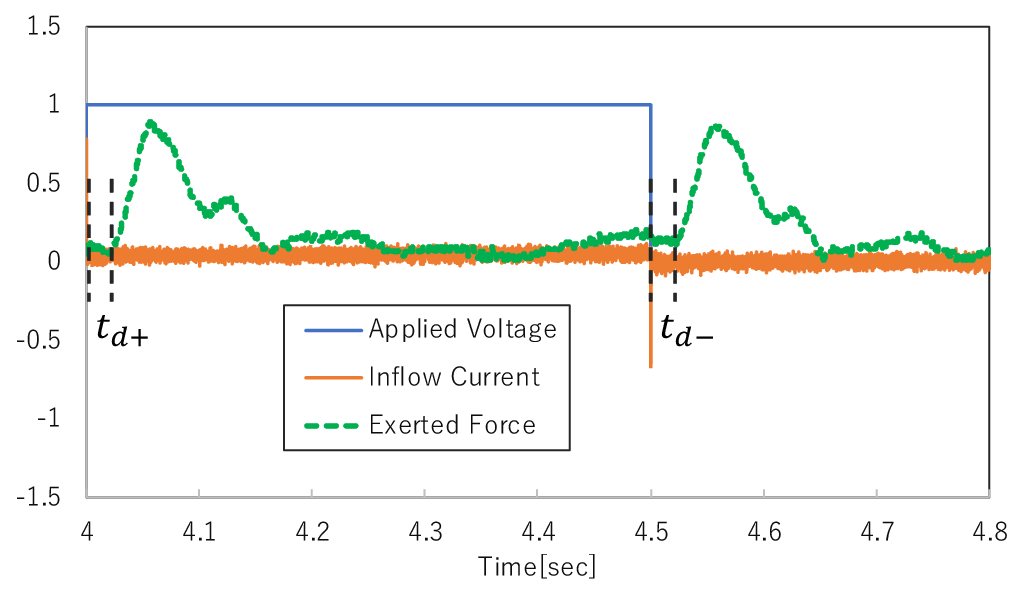}
    \caption{Applied voltage, inflow current, and exerted force (normalized to a maximum value of one) after stimulation of flexor muscle of subject A}
    \label{fig:Dead} 
\end{figure}

Next, we confirmed which of the applied voltages and the inflow currents are related to the exerted force. The subjects were given electrical stimulation with a pulse width of 500 msec and amplitude of 20 V. Fig.~\ref{fig:Dead} shows the results of normalizing the maximum value, inflow current, and exerted force of the applied voltage of subject A with 1. Three points can be confirmed from the result.
\begin{itemize}
  \item A dead time of about 20 msec passed until the force was generated after the flow of current.
  \item The exerted force was generated with a short dead time after the flow of current. In addition, no current flowed for about 100-400 msec, and no force is generated at the time of voltage application. Therefore, the force was determined by the inflow current.
  \item Forces were generated by the positive current and the negative current, and the characteristics forces were different from each other.
\end{itemize}

The dead time of each subject was estimated from this experiment. The average of five experimental results was defined as the dead time of each subject. Table~\ref{tb:Value2} shows the dead times of each subject. The subscripts + and -- denote positive and negative currents.

\begin{table}[tbh]
\centering
\caption{Values of Dead Time $t_{d}$}
\label{tb:Value2}
  \begin{tabular}{|c||c|c|} \hline
     & Subject A & Subject B \\ \hline\hline
    $t_{d+}[sec]$ & 0.023 & 0.021  \\ \hline
    $t_{d-}[sec]$ & 0.025 & 0.028 \\ \hline
  \end{tabular}
\end{table}

\subsection{Measurement of Threshold Current}
\begin{figure}[t]
\centering
\includegraphics[width=80mm]{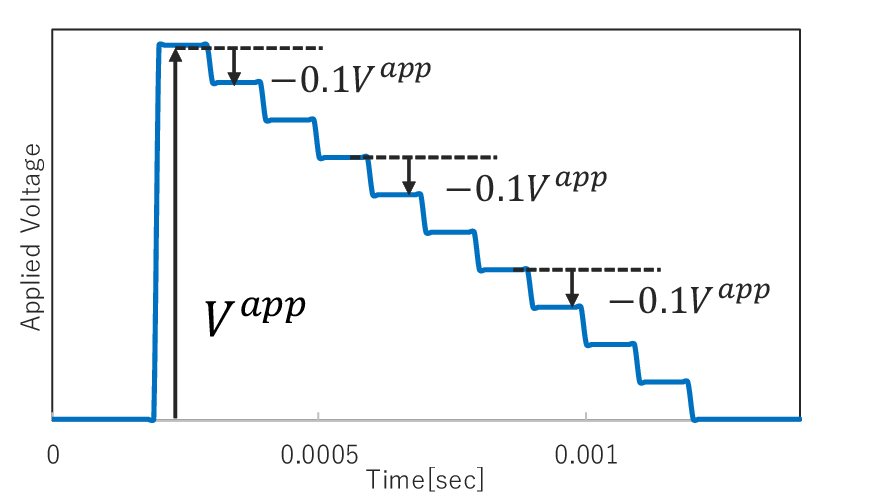}
    \caption{Shape of stimulation waveform that can check the effect of only one side current}
    \label{fig:Waveform} 
\end{figure}

\begin{figure}[t]
\centering
\includegraphics[width=90mm]{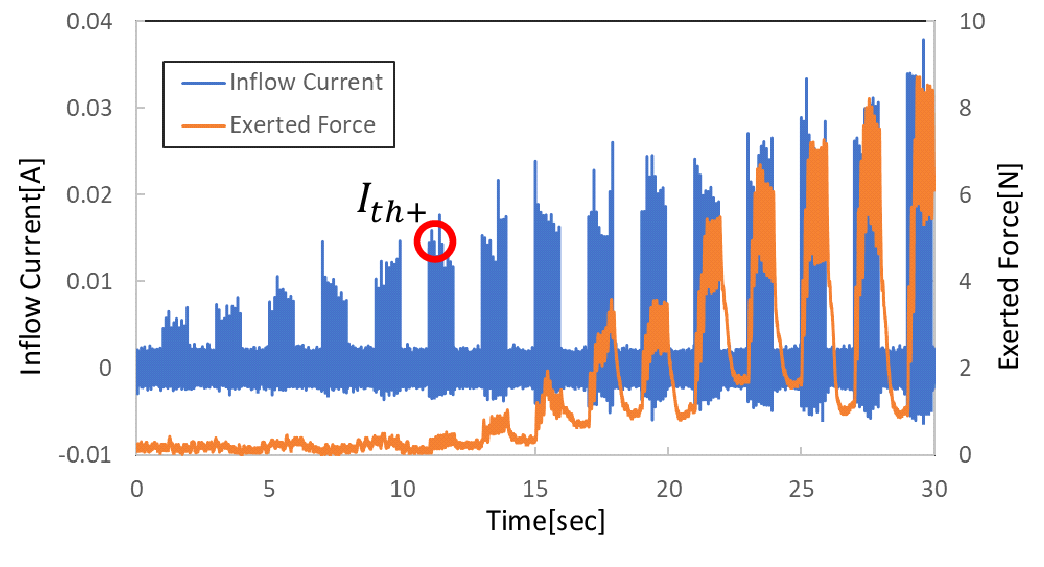}
    \caption{Experimental results of temporal change in electric current and exerted force of subject B}
    \label{fig:Ith} 
\end{figure}

Next, we gradually increased the current flowing into the subject and measured the threshold current required to drive the joint. In the experiment, the voltage corresponding to the stimulation waveform shown in Fig.~\ref{fig:Waveform} was applied. The reason for using this shape is that the negative current can be reduced by the waveform shown in Fig.~\ref{fig:Waveform}, and only the force generated by the positive current can be measured. In addition, by exchanging the positive and negative sides of the stimulation waveform of Fig.~\ref{fig:Waveform}, only the force due to the negative current can be measured. The stimulation frequency was set to 10 pps (pulses per second), stimulation was performed for 1 sec, and a 1 sec break was set. The initial voltage was 6 V, and it was increased in steps of 2 V. The results of subject B are shown in Fig.~\ref{fig:Ith}. There exists a threshold current at which the subject's joints exert their forces. When the direction of the current changed, the same result was obtained in the experiment, even with a different subject. The threshold current $I_{th}$ values of each subject are listed in Table~\ref{tb:Value3}.

\begin{table}[tbh]
\centering
\caption{Values of Threshold Current $t_{d}$}
\label{tb:Value3}
  \begin{tabular}{|c||c|c|} \hline
     & Subject A & Subject B \\ \hline\hline
    $I_{th+}[mA]$ & 14.4 & 15.1 \\ \hline
    $I_{th-}[mA]$ & 8.32 & 12.3 \\ \hline
  \end{tabular}
\end{table}

\subsection{Relationship between Inflow Current and Exert Force}
For the two subjects, an experiment was conducted in which the stimulation waveform shown in Fig.~\ref{fig:Waveform} was randomly applied at intervals of 1 msec. The timing of applying the stimulus was 1000 pps, but the probability of applying the stimulus was 1/2. Therefore, a pseudo M-sequence signal was generated. The maximum amplitude of the voltage was 20 V for both subjects. The inflow current $I ^ {flo}$ and the exerted force $f ^ {ext}$ were acquired with a sampling period of 0.1 msec. There after that, decimation was performed to limit the sampling period to 5.0 msec. We estimated the transfer function $ G_ {IF} (s) $ using the method described in Section III with the 512 pieces of data obtained. The current $I$ used in the verification is expressed as follows:
\begin{equation}
I=
\begin{cases}
\; I^{flo}-I_{th} &$($I^{flo} > I_{th}$) $\\
\; 0 &$($I^{flo} \le  I_{th}$)$.
\end{cases}
\label{eq:I}  
\end{equation}

\begin{figure}[t]
\begin{center}
\includegraphics[width=80mm]{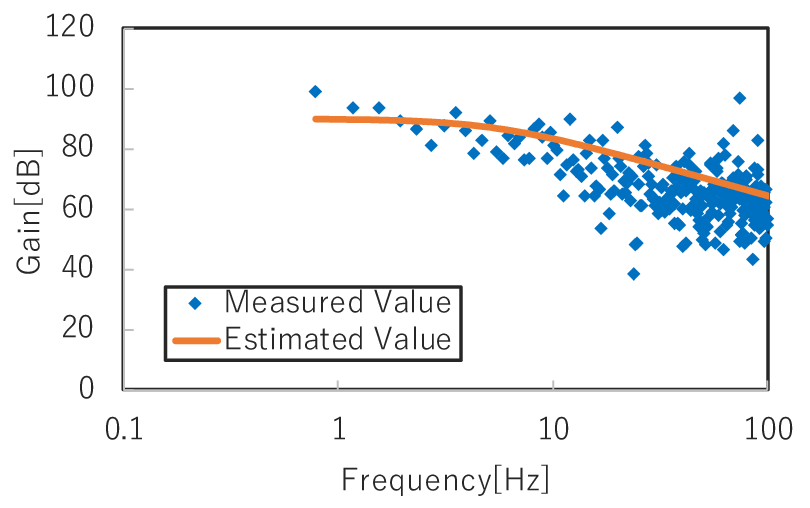}
\caption{Gain margin of positive current of subject A}
\label{fig:GIF1}
\end{center}
\begin{center}
\includegraphics[width=80mm]{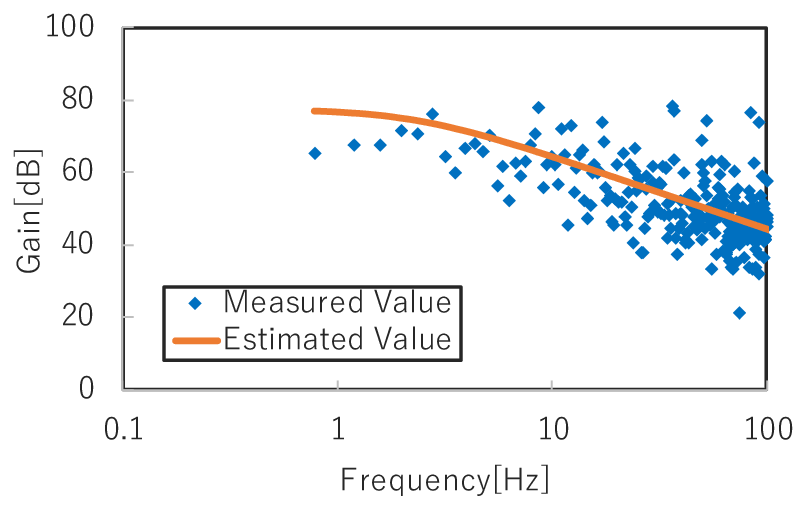}
\caption{Gain margin of negative current of subject B}
\label{fig:GIF2}
\end{center}
\end{figure}

The gain margin in the case of a positive current applied to subject A and the gain margin in the case of a negative current applied to subject B are shown in Figs.~\ref{fig:GIF1},~and~\ref{fig:GIF2}, respectively. The results shows that the gain margin changes from a certain point at --20 dB/dec. Therefore, it is inferred that the inflow current and the exerted force constitute a first-order lag system. The transfer function $ G_ {IF +} (s) $ is written with the constants $ c_{1}, d_{0} $:
\begin{equation}
 G_ {IF+} (s)=\frac{F_{+}(s)}{I_{+}(s)}=\frac{d_{0}}{c_{1}s+1},
\label{eq:TF_IF}  
\end{equation}

These two values were estimated using multiple regression analysis. The fixed values of each subject are shown in Table~\ref{tb:Value4}. 

\begin{table}[tbh]
\centering
\caption{Values of parameter concerning $ G_ {IF} (s) $}
\label{tb:Value4}
  \begin{tabular}{|c||c|c|} \hline
     & Subject A & Subject B \\ \hline\hline
    $c_{1+}$ & 0.1889 & 0.5789  \\ \hline
    $d_{0+}$ & 32207 & 4888.2 \\ \hline
    $c_{1-}$ & 0.2476 & 0.4325 \\ \hline
    $d_{0-}$ & 13796 & 7331.5 \\ \hline
  \end{tabular}
\end{table}

The outline of the transfer function obtained using the values listed in Table~\ref{tb:Value4} and (\ref{eq:TF_IF}) is shown by the solid line in Figs.~\ref{fig:GIF1},~and~\ref{fig:GIF2}. It was confirmed that the relationship between the inflow current and the exerted force agrees with the result obtained using FFT. From the results, it can be inferred that the relationship between the inflow current and the exerted force is a first-order lag system.

\section{Verification}
\begin{figure}[t]
\centering
\includegraphics[width=90mm]{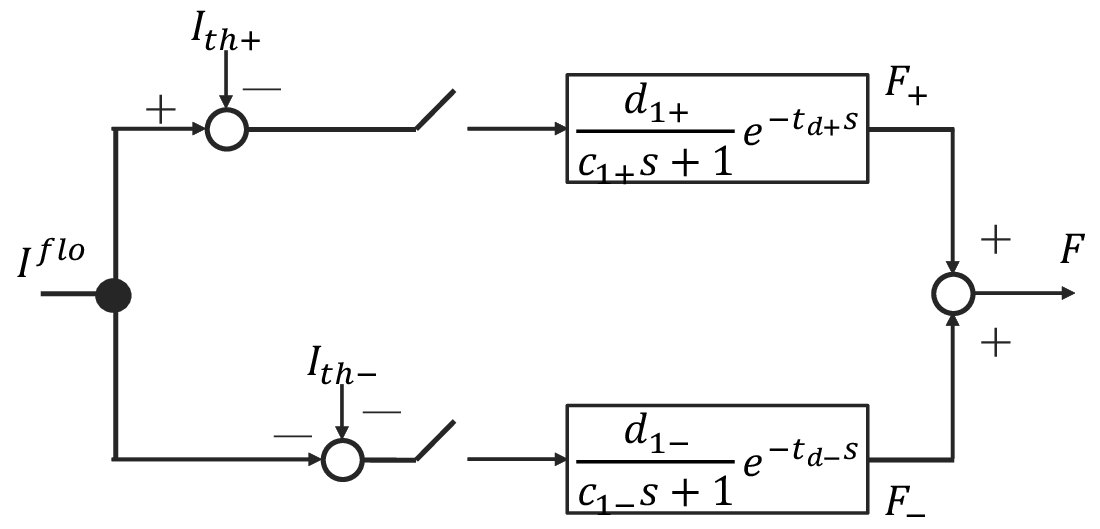}
    \caption{The structure of model of human muscle}
    \label{fig:Block} 
\end{figure}

In this section, we verify the accuracy of the force-of-force estimation by employing a stimulation pattern different from the one used in the previous section. From the results in Fig.~\ref{fig:Dead}, the positive current and negative current exert power independently. Therefore, the model of human muscle (i.e, the relationship between current and force) is defined in Fig.~\ref{fig:Block}. The stimulus pattern used for verification was an M-sequence signal that changed at +15 V and --15 V at 1000 pps. 

\begin{figure}[t]
\begin{center}
\includegraphics[width=80mm]{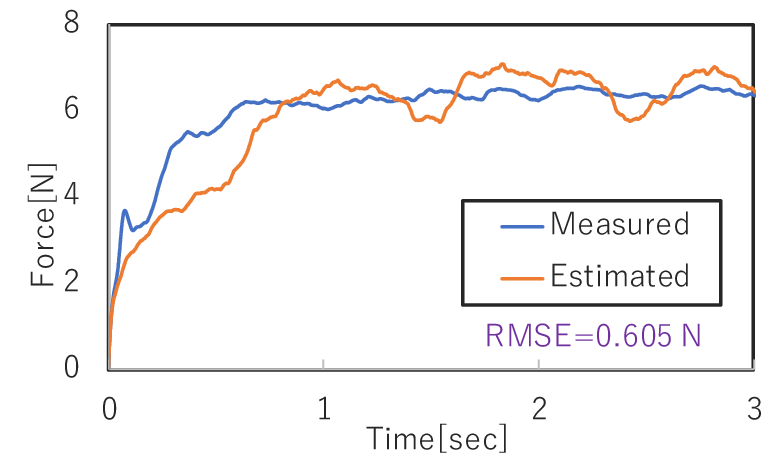}
\caption{Actual measured value and estimated value of subject A}
\label{fig:OutputA}
\end{center}
\begin{center}
\includegraphics[width=80mm]{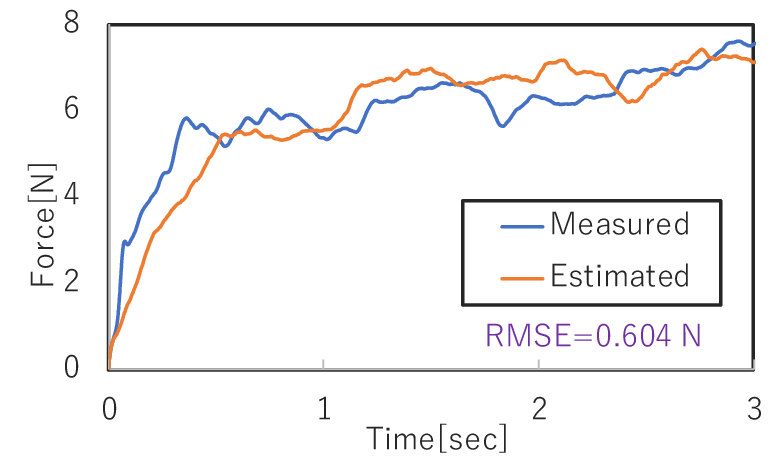}
\caption{Actual measured value and estimated value of subject B}
\label{fig:OutputB}
\end{center}
\end{figure}

The measured and estimated values of the exerted force are shown in Figs.~\ref{fig:OutputA},~and~\ref{fig:OutputB}. In addition, the current and the exerted force are passed through a low-pass filter with a cutoff frequency of 100 Hz.

From the results, it is clear that the error is large from the start to about 0.5 sec, but thereafter, the two values agree reasonably. One of the factors that caused errors after starting was incorrect modeling of the transient response. Because the data used to create the model in the proposed method were related to stationary response, there is a high possibility that the model cannot handle transient responses. The goal is to improve the estimation accuracy for transient response in the future.

\section{Conclusion}
In this paper, we proposed a method modeling the relationship between current and exerted force when using FES. First, we showed that the exerted force is determined by the inflow current. Next, we found that the relationship between the inflow current and exerted force can be formulated as a first-order lag system including dead time. As a result, we confirmed that the proposed method can be used to estimate the ability to exerted force under steady state response. In addition, this result suggests that the method using the conventional pulse width modulation (PWM) method \cite{c23} \cite{c24} can be applied. By adopting PWM method, performance improvement of control using FES is expected. In the future, we will improve the transient response of the model and improve the overall musculoskeletal system.


\section*{Acknowledgement}
This work was entrusted by the KDDI Foundation and JST, PRESTO Grant Number JPMJPR1755, Japan.


%

\end{document}